\documentclass[aps,pre,preprint,onecolumn]{revtex4-1}
\usepackage{amsmath}
\usepackage{graphicx}

\begin{document}

\title{Bubble Collapse and Jet Formation in Corner Geometries}

\author{Yoshiyuki Tagawa$^1$}
\author{Ivo R. Peters$^2$}
\email[]{i.r.peters@soton.ac.uk}
\affiliation{
$^1$Department of Mechanical Systems Engineering, Tokyo University of Agriculture and Technology, Tokyo, Japan\\
$^2$Engineering and the Environment, University of Southampton, Highfield, Southampton SO17 1BJ, UK}

\date{\today}

\begin{abstract}
The collapse of a vapor bubble near a flat solid boundary results in the formation of a jet that is directed towards the boundary. In more complex geometries such as corners, predictions of the collapse cannot be made in a straightforward manner due to the loss of axial symmetry. We experimentally investigate the bubble collapse and jet formation in corners formed of two flat solid boundaries with different opening angles. Using potential flow analysis, we accurately predict the direction of the jet and bubble displacement. We further show that for a corner with an opening angle $\alpha$, there exist analytic solutions that predict the jet direction for all the cases $\alpha=\pi/n$, where $n$ is a natural number. These solutions cover, in discrete steps, the full range of corners from the limiting case of a bubble near a single wall ($n=1$) up to a bubble in between parallel walls ($n\rightarrow\infty$).
\end{abstract}

\maketitle

%%
%%
%%
%% 1. INTRODUCTION
%%
%%
%%

The growth and collapse of vapor bubbles has been investigated extensively since the early works of Rayleigh \cite{Rayleigh1917}. One  aspect of the collapse of these bubbles is the jet that forms near a rigid boundary, a phenomenon of specific interest for cavitation damage, ultrasonic cleaning, and numerous medical applications such as root canal cleaning or directed drug delivery through sonoporation \cite{Benjamin1966, Plesset1971, Blake1987, Philipp1998, Unger2004, Bremond2006, Ohl2006, VanderSluis2007, DeGroot2009, Vos2011}. This non-spherical collapse near a wall has been treated in many experimental and numerical studies, which have focused on, among others, the evolution of the bubble shape, jet velocity, impact stress, and cleaning capabilities \cite{Vogel1989, Dijkink2008, Johnsen2009}. Although complexity has been gradually increased over the years by adding more bubbles or investigating cases of curved or soft boundaries, these studies remain in a domain of axial symmetry \cite{Blake1993, Tomita2002}. A much more extensively investigation of the influence of geometry has been done in the highly confined systems of microfluidic devices, where flow is approximated as either one-dimensional or two-dimensional \cite{Chen2006, Zwaan2007, Sun2009, Tagawa2012b}. In large, three-dimensional systems, however, surprisingly few investigations have been done in geometries that break the axial symmetry. A notable exception to this is the recent experimental study on cavitation near two perpendicular walls by Ref~\cite{Brujan2018}.

The non-spherical collapse of a single bubble is a highly non-linear process. The large surface deformations severly limits the use of analytical descriptions, and detailed information therefore needs to be obtained by numerical simulations or experiments \cite{Blake1987, Johnsen2009}. Qualitatively, however, the initiation of the non-spherical collapse near a wall can be described in a straightforward manner using potential flow theory. By representing the bubble with a sink and modeling the wall with an image sink, the induced velocity of the image sink on the bubble gives the direction of the jet \cite{Plesset1977, Tomita2002}. The displacement of the bubble can be further quantified by using the flow potential prescribed by the two sinks to calculate the Kelvin impulse \cite{Blake1988,Supponen2016}, but this requires using the approximation that the bubble remains spherical during the collapse. Here, we will investigate the case of two walls that form a corner with an opening angle $\alpha$, and determine how the direction of the jet that forms during the collapse depends on $\alpha$, and the position and size of the bubble. We will present experimental results and introduce a model that provides analytic solutions for a family of corners. We confirm the accuracy of our predictions by direct comparison with our experimental results.

%%
%%
%%
%% 2. EXPERIMENTAL
%%
%%
%%

\begin{figure*}
	\centering
	\includegraphics{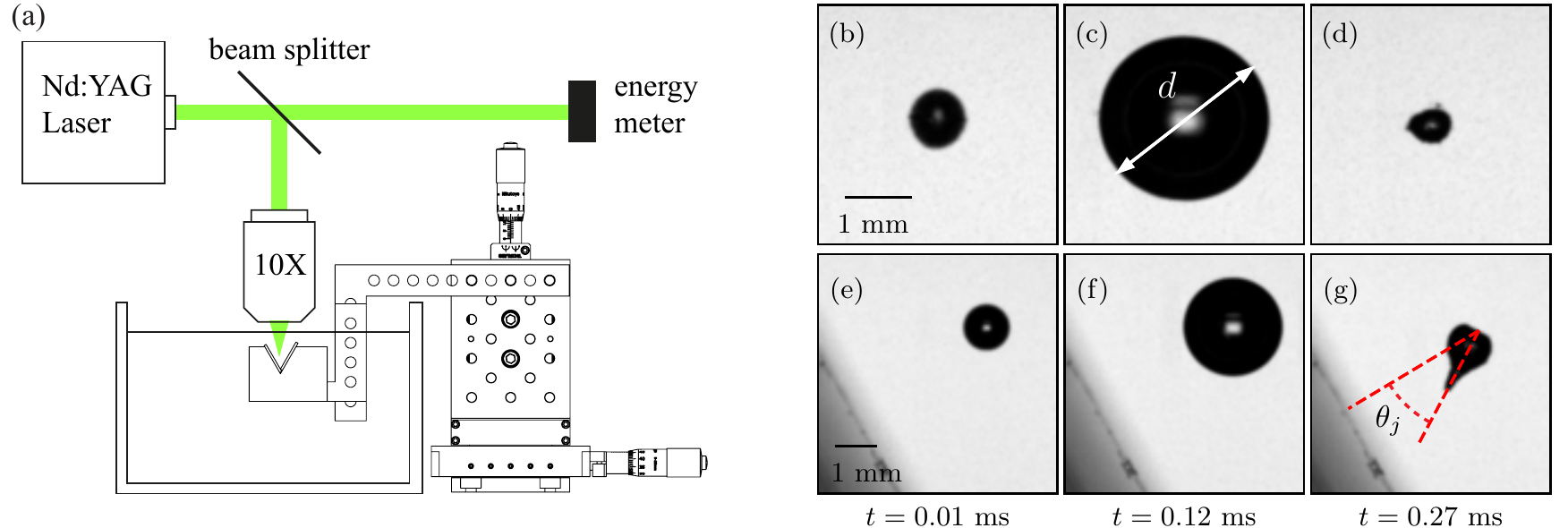}
	\caption{Schematic drawing of the experimental setup, excluding the camera, lens, and light source for imaging (a) and snapshots from high-speed recordings (b-g). Growth and collapse of a free bubble (b-d) and a bubble in a corner with opening angle $\alpha=\pi/3~\mathrm{rad}$ (e-g). Both bubbles were created by a laser pulse at $t=0$. The free bubble shows an axisymmetric collapse and remains in its original position, while the bubble in the corner generates a jet and is displaced toward the direction of the jet. We define the jet direction $\theta_j$ with respect to the normal of one of the solid surfaces.}
	\label{fig:fig1}
\end{figure*}

In our experiments, we produced vapor bubbles using laser-induced cavitation \cite{Brewer1964, Felix1971, Vogel1996}. We focused a pulsed laser inside a large water bath ($180\times180\times100~\rm mm^3$) and recorded the growth and collapse of the bubbles using a high-speed camera. Figure \ref{fig:fig1}a shows a schematic view of the setup. A Q-switched Nd:YAG laser (Bernoulli PIV, Litron) provides pulses with a duration of 6 ns and an energy up to 200 mJ per pulse at a wavelength of 532 nm. We varied the energy per pulse between 10 and 18 mJ by adjusting either the driver power or attenuation of the laser. Using a 50:50 beamsplitter we passed half of the pulse to an energy meter. The other half of the pulse was passed through a $10\times$ microscope objective (Nikon Plan Fluor, \emph{NA}=0.30) positioned $\sim 1~\mathrm{mm}$ above a water surface as illustrated in Fig. \ref{fig:fig1}a. The laser was focused at a distance $h\approx{20~\mathrm{mm}}$ below the water surface. With a maximum bubble diameter $d=3.3~\mathrm{mm}$ the stand-off distance, $h/d\gtrsim6$, was large enough to neglect the influence of the free surface on the collapse \cite{Blake1987b}. The target geometry consisted of two microscope slides ($25.8\times 75.7\times 1.4~\mathrm{mm^3}$) mounted in a holder at an opening angle $\alpha$ of either $\pi/3$ or $\pi/2~\mathrm{rad}$. The relative position of the microscope slides with respect to the focal point of the microscope objective was controlled within $5~\mathrm{\mu m}$ using an XYZ translation stage. Experiments were recorded using a high-speed camera (Photron FASTCAM SA-X2) at an acquisition rate of 100 kHz, generating images of $384\times 264$ pixels. The exposure time was set to $1~\rm\mu s$ to minimize motion blur. We used a 105 mm Nikon Micro-Nikkor lens, resulting in a resolution of 21.2 to 31.3 $\rm\mu m/\rm pixel$. A 550 nm longpass filter was positioned in front of the camera lens to prevent any laser light to enter the imaging system.

Calibration of the absolute distance between the focal point of the laser and the microscope slides was done by generating a vapor bubble near one of the slides and determining the position of the glass surface from the reflection of the jet that was formed during the collapse. We used these reference positions in combination with the displacement measured with the translation stage to obtain a mapping of pixel positions to physical distances when the microscope slides were out of view. The uncertainty in this procedure is of the order of 2 pixels, leading to a total uncertainty of 47 to 67 $\mu m$ in the position with respect to the walls, depending on the resolution of a particular high-speed movie.

Experiments were run by using the high-speed camera as a master and using the trigger output of the camera to trigger the laser. With this synchronized setup, a single laser pulse was generated $156~\rm\mu s$ after the recording started, \emph{i.e.}, $6~\rm\mu s$ after the 16$^{\rm th}$ frame at our acquisition rate. 

\begin{figure}%[htbp]
	\includegraphics{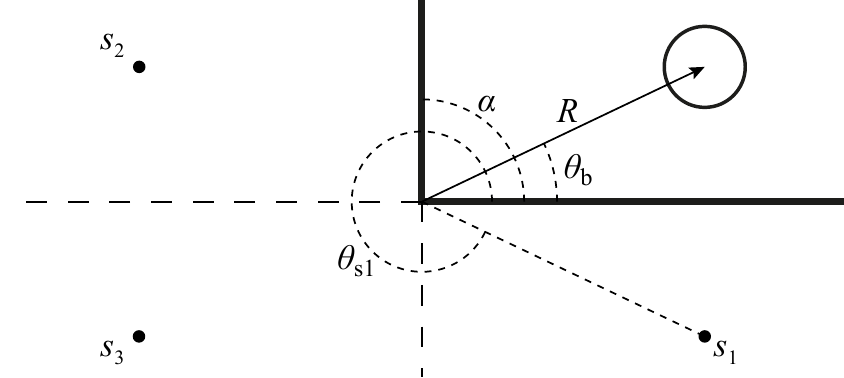}
	\caption{Schematic representation of a bubble at a position $\vec{r}_j=Re^{i\theta_b}$ in a corner with an opening angle $\alpha=\pi/2$. The walls (thick solid lines) are modeled using three image sinks $s_{1-3}$ which together induce a velocity $\vec{u}_j$ at the bubble position. The bubble center and the sink $s_1$ are at the same vertical distances from the horizontal wall. The same symmetrical condition is applied to the bubble center, sink $s_2$, and the vertical wall. Radial distance from the corner is $R$ for all sinks.}
	\label{fig:model}
\end{figure}

\begin{figure}%[htbp]
	\includegraphics{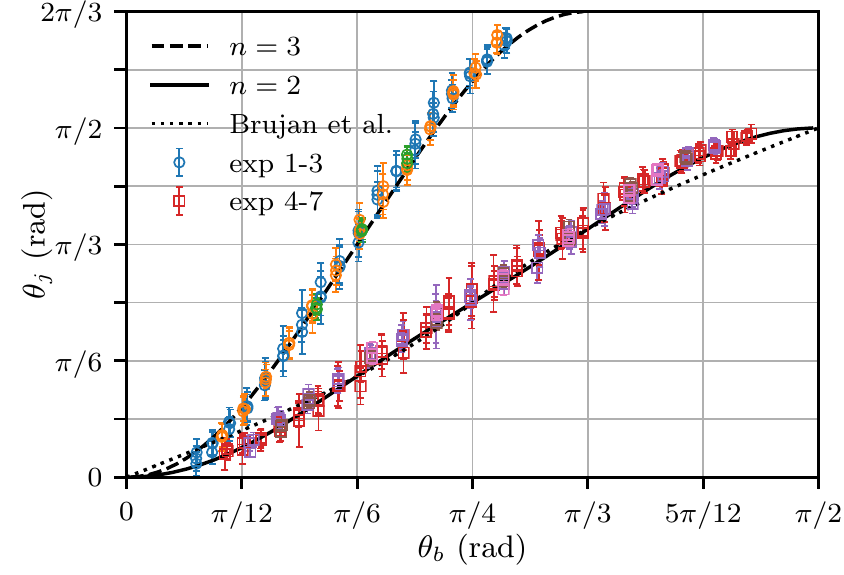}
	\caption{Comparison between experimentally determined jet angles (symbols) and model predictions (solid and dashed lines) without any adjustable parameters. Experiments 1-3 are performed at various distances from the $\pi/3~\mathrm{rad}$ corner, experiments 4-7 are performed in a $\pi/2~\mathrm{rad}$ corner at various distances as well as different bubble sizes. The correlation found by Brujan et al. \cite{Brujan2018} is shown by the dotted line.}
	\label{fig:expModelComparison}
\end{figure}

We obtained the parameters bubble size, bubble position and jet direction, from our experiments using image processing. All images were processed with a standard procedure where background subtraction, thresholding, removing noise and filling holes were applied to obtain clean binary images where the bubble is fully converted into black pixels. This allowed us to directly obtain the two-dimensional area (number of pixels) and the position of the bubble center in the image as a function of time. From the image where the bubble has reached its maximum size, we determined the bubble position $(R, \theta_b)$ as defined in Fig.~\ref{fig:model} and the diameter $d$ (see Fig. \ref{fig:fig1}c). We chose this point as a reference for the position and the size because the bubble is at its most spherical shape, which results in an unambiguous determination of the position and the size. The collapse following from this point makes the bubble increasingly non-spherical and eventually results in the formation of the jet. Instead of measuring the direction of the jet directly, we determined the displacement of the bubble during the first collapse and rebound. For this we obtained the position of the bubble using the same method as before, with the only difference being the use of the image where the bubble size has reached its second maximum. The absolute position based on the second maximum could not be determined accurately because the bubble in this stage is far from spherical, as can be observed in Fig. \ref{fig:fig1}g. For the determination of the direction of displacement, however, this non-spherical shape is of little influence because the bubble is symmetric around the direction of displacement as can be seen in the same figure. From these displacements we then calculated the angle of displacement $\theta_j$ with respect to the normal to one of the walls as indicated in Fig.~\ref{fig:fig1}g. We took into account an uncertainty in the determination of the bubble positions of 1 pixel. The resulting uncertainty in the angle, $\Delta\theta_j$, is represented by the error bars in Fig. \ref{fig:expModelComparison}. To verify our method of determining the direction of the jet, we compared the angle measured from the displacement and the directly measured angle for a sample of our experiments and found a typical difference of 0.02 rad, well within the measurement uncertainty and the spread in the data.

%%
%%
%%
%% 2b. MODEL
%%
%%
%%

We model our experiments as incompressible and inviscid flow, such that we can use the potential flow approximation where the velocity potential $\phi$ satisfies the Laplace equation $\nabla^2\phi=0$. This requires that viscous effects are confined to thin boundary layers at the solid surfaces. From the typical timescale of $t_0\sim0.3~\mathrm{ms}$ (Fig.~\ref{fig:fig1}b-g), we estimate the maximum boundary layer thickness to be of the order of $\delta\sim\sqrt{\nu t_0}\approx 17~\mathrm{\mu m}$, where $\nu$ is the kinematic viscosity of water, a length scale much smaller than the typical length scale of our experiments. Compressibility effects can be neglected during most of the bubble growth and collapse, except at times close (within $\sim 10 ~\mathrm{\mu s}$) to bubble formation and final stage of collapse when shock waves are formed \cite{Tagawa2016}. We are however interested in the collapse dynamics right after the bubble has reached its maximum diameter ($t\approx 0.12~\mathrm{ms}$), which is when speeds are minimal.

Our model consists of an infinite three-dimensional domain, where the collapsing bubble is replaced by a three-dimensional sink and the walls are modeled by placing image sinks. To adhere to no-penetration boundary conditions at both walls, we introduce, for instance for a corner angle $\alpha=\pi/2$, three image sinks as indicated in Fig. \ref{fig:model}. Note that a superposition of the solutions of two individual walls (i.e., only keeping sinks $s_1$ and $s_2$) does not result in a solution that satisfies the no-penetration boundary conditions on the walls. Using radial flow and spherical symmetry, each sink induces a velocity field described as $u_r=-m/(4\pi r^2)$, with $m$ the sink strength (flow rate) and $r$ the distance from the sink.
To predict the direction of the jet resulting from the collapse of the bubble, we calculate the velocity at the bubble position that is induced by the image sinks. The contribution of the sink that represents the bubble itself is discarded because it does not contribute to a net velocity at the bubble position due to spherical symmetry. Here we make the assumption that the direction of the induced velocity vector coincides with the direction of the jet. This is true for the collapse of a bubble near a single solid wall, but we show here that this also holds for corner geometries.

The velocity vector $\vec{u}_s$ at the bubble position $\vec{r}_b=Re^{i\theta_b}$, induced by a single sink at position $\vec{r}_s=Re^{i\theta_s}$ is obtained by taking the magnitude of the sink flow $|\vec{u}_s|=m/(4\pi r^2)=m/(8\pi R^2[\cos(\theta_b-\theta_s)])$, and multiplying this with the unit vector $\hat{r}_s$ that points to the position of the sink with respect to the bubble \mbox{$\hat{r}_s=(e^{i\theta_s}-e^{i\theta_b})/(\sqrt{2[1-\cos(\theta_b-\theta_s)]})$}. This product results in the general expression
\begin{equation}
	\vec{u}_s=\frac{m}{\pi\sqrt{128}R^2}
	\frac{e^{i\theta_s}-e^{i\theta_b}}{[1-\cos(\theta_b-\theta_s)]^{3/2}},
\end{equation}
which can be used to calculate the induced velocity for any image sink. Here, we have made use of the fact that all image sinks are at the same distance $R$ from the corner as the bubble. For the example above where $\alpha=\pi/2$ we would have three image sinks, and consequently three values of $\vec{u}_s$ that would sum up to the predicted jet direction given by $\vec{u}_j$. We will now proceed to show that the analysis can be generalized to a corner angle $\alpha=\pi/n$, where $n$ is a natural number. The number of image sinks will be $2n-1$, with one sink at an angle $(2\pi-\theta_b)$, $n-1$ sinks at $(2\pi k/n-\theta_b)$, and $n-1$ sinks at $(2\pi k/n+\theta_b)$, where $k$ indicates the index running from 1 to $n-1$.

Summing over $2n-1$ image sinks in a corner $\alpha=\pi/n$ then results in
\begin{equation}
	\label{eq:model}
	\frac{\vec{u}_j}{C} =
	\frac{e^{i\theta_b}-e^{i(2\pi-\theta_b)}}
	{[1-\cos(2\theta_b)]^{3/2}} +
	\sum\limits_{k=1}^{n-1}\left\lbrace
	\frac{e^{i\theta_b}-e^{i(2\pi k/n-\theta_b)}}
	{[1-\cos(2\theta_b-2\pi k/n)]^{3/2}}\right. +
	\left.\frac{e^{i\theta_b}-e^{i(2\pi k/n+\theta_b)}}
	{[1-\cos(2\pi k/n)]^{3/2}}\right\rbrace,
\end{equation}
with the prefactor $C=m/(\pi\sqrt{128}R^2)$. The jet direction is then given by $\cos\theta_j=\mathrm{Im}(\vec{u}_j)/|\vec{u}_j|$, which removes the prefactor $C$ from the solution. Note that for $n=1$ only the first term remains, recovering a jet pointed towards a single wall. 

%%
%%
%%
%% 3. RESULTS
%%
%%
%%

\begin{figure}%[htbp]
	\includegraphics{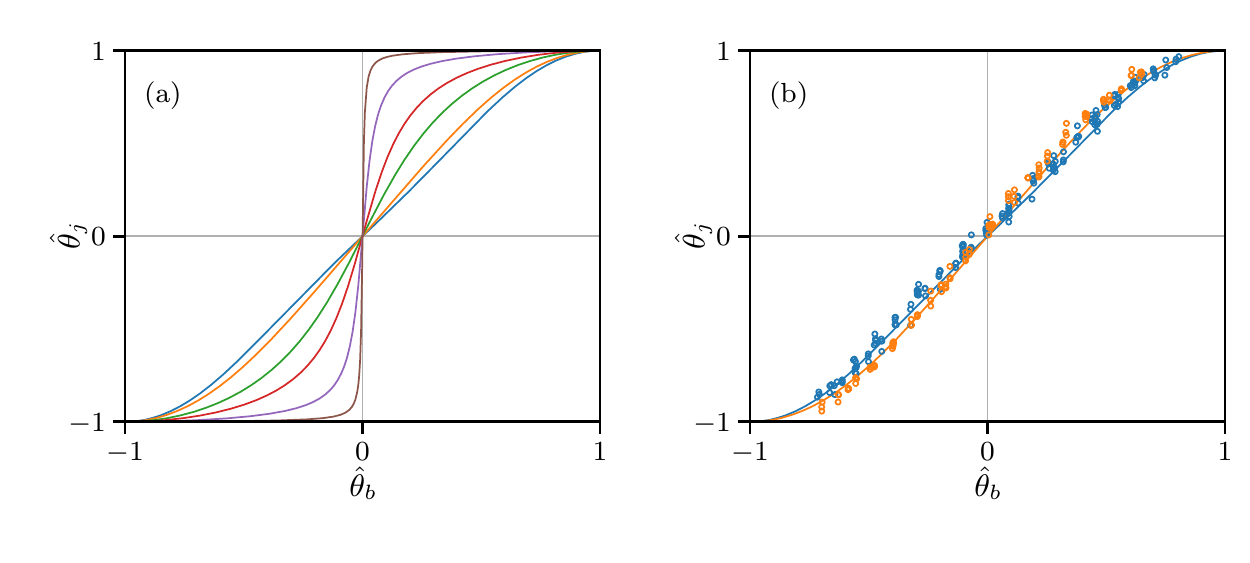}
	\caption{(a) Normalized model results for $n=2$ (blue), $n=3$ (orange), $n=8$ (green), $n=20$ (red), $n=100$ (purple), and $n=1000$ (brown). The curve approaches a step function as $n\rightarrow\infty$. (b) Normalized experimental data (symbols) and model (lines) for $n=2$ (blue) and $n=3$ (orange).}
	\label{fig:curve_collapsing}
\end{figure}

Our results are summarized in Fig. \ref{fig:expModelComparison} and \ref{fig:curve_collapsing}. Our main experimental result is the dependence of the jet direction as a function of the initial angular position of the bubble. The measurements were performed by making horizontal sweeps to obtain a range of values for $\theta_b$. We repeated these sweeps at various vertical positions to test for dependence on absolute distance. From the collapse of the data in Fig. \ref{fig:expModelComparison} it is clear that the radial distance from the corner to the bubble has no influence on the direction, and it is only the angular position that determines the outcome. We have additionally varied the laser energy to obtain different bubble sizes ($d=2.3$ to $3.3~\mathrm{mm}$), and show that the angle is also independent of the diameter $d$ of the bubble, as predicted. The angles of the jets that we obtain nearly cover the full possible range of $\theta_j=0$ to $\theta_j=\pi/2$ rad and $\theta_j=2\pi/3$ rad, for $\alpha=\pi/2$ rad and $\alpha=\pi/3$ rad, respectively. 

We compare the data in Fig.~\ref{fig:expModelComparison} with our model prediction (\ref{eq:model}) based on 3D sinks. Both the $\pi/2$ rad corner (solid line) and the $2\pi/3$ rad corner (dashed line) show an excellent quantitative agreement with our experimental data without any adjustable parameters. The recently reported empirical relation $\theta_j\propto x_0/y_0$ for the jet direction of a collapsing bubble near two perpendicular walls is shown by the dotted line \cite{Brujan2018}. We have mirrored this correlation around $\theta_b=\pi/4~\mathrm{rad}$ to compare the relation to our full data set. Although it shows a good approximation to our experimental data and the sink model near $\theta_b=\pi/4$, our model provides better predictions for the approach to the limiting angles $\theta_b=0$ and $\theta_b=\pi/2$. In particular, our model shows the nonlinear approach to the limiting angles, characterized by the slope $d\theta_j/d\theta_b=0$, which is a general feature for all opening angles and a result of the three-dimensional sink flow \footnote{The same model can be constructed with 2D sinks, where $u_r=-m/(2\pi r)$, which results in $\theta_j\propto\theta_b$ in the limits $\theta_b\rightarrow 0$ and $\theta_b\rightarrow \alpha$.}.

As the initial bubble position approaches a wall, the jet direction becomes perpendicular to that wall. This means that the smaller the opening angle $\alpha$ of the corner (a smaller range in $\theta_b$), the larger the range in jet angles. Defining the jet angle $\theta_j=0$ at $\theta_b=0$, the maximum expected jet angle will be $\pi-\alpha$, as confirmed in Fig. \ref{fig:expModelComparison}. We explore this common behavior by using the corner opening angle to obtain a normalized bubble position $\hat\theta_b=2\theta_b/\alpha-1$ and normalized jet direction $\hat\theta_j=2\theta_j/(\pi-\alpha)-1$, where the hat indicates the normalized version of the angle, so that both values will range from $-1$ to $1$. In Fig. \ref{fig:curve_collapsing}(a) we show a set of results from our model in these normalized angles. All curves have a smooth sigmoid-like shape with a vanishing slope $d\hat\theta_j/d\hat\theta_b\rightarrow 0$ as $\hat\theta_b\rightarrow -1$ and $\hat\theta_b\rightarrow 1$. The slope at $\hat\theta_b=0$ increases with $n$, and the curve approaches a step-function as $n\rightarrow\infty$ (or $\alpha\rightarrow 0$). This limit corresponds to a bubble between two parallel plates, where, if a jet is formed, the jet points towards the nearest wall and not in any other direction due to symmetry. This behavior is however only expected when the size of the bubble is much smaller than the gap width between the parallel plates, otherwise different collapse scenarios can be observed \cite{Gonzalez-Avila2011a,Quah2017}

Figure \ref{fig:curve_collapsing}(b) shows our experimental data for the two opening angles in the normalized form. The normalization brings these curves very close together, but still the curves can be distinguished and follow the predictions of their specific opening angle. Cases where bigger differences are expected (\emph{e.g.}, the curves for $n=8$ or $n=20$ become increasingly challenging to test experimentally due to the small opening angle of the corner. This would require bubble sizes significantly smaller than the distance between the solid boundaries or an increased size of the experimental setup.

%%
%%
%%
%% 4. CONCLUSIONS & DISCUSSION
%%
%%
%%

We have shown that the direction of a jet resulting from the collapse of a vapor bubble in a corner with an opening angle $\alpha=\pi/n$ can be fully described using potential flow theory with a finite distribution of three-dimensional sinks. The analytic results accurately predict the jet directions measured experimentally. 
Although the collapse mechanism is highly nonlinear, the remarkable quantitative agreement indicates that the flow during the initial stage of the collapse is the dominant factor in determining the direction of deformation and the eventual direction of the jet. This analysis could be further refined and be used to predict the jet strength by calculating the Kelvin impulse and taking into account the time-dependence of the sinks \cite{Supponen2016}.
We have further shown that the normalized predictions all produce a smooth step function. Our analytic solutions could be used to predict the jet direction for corners with intermediate angles to the ones presented here by interpolation the solutions. Considering that numerical simulations of the collapse of vapor bubbles are computationally expensive, our results can be used to make analytic predictions for bubble migration and cavitation damage sites or to predict the effectiveness of ultrasonic cleaning. Such predictions can be utilized in the design of components that need to be protected from cavitiation. Further studies will be needed in more complex geometries or with multiple bubbles, where our findings can provide a starting point.

\begin{acknowledgments}
We thank D. Fern{\'a}ndez Rivas, D. Lasagna, D. van der Meer, and C.-D. Ohl for insightful discussions, and A. Franco-G{\'o}mez for proofreading the manuscript. We thank the Daiwa Foundation (No. 8590/12337) for financial support, YT acknowledges financial support from JSPS KAKENHI grant number 17H01246, and IRP acknowledges financial support from the EPSRC under grant number EP/P012981/1
\end{acknowledgments}

\end{document}